\documentclass[useAMS,usenatbib,usegraphicx]{mn2e}

\title[Transit timing variation in exoplanet WASP-3b]{Transit timing variation in exoplanet WASP-3b\thanks{This work is based on observations made with the 60-cm telescope of the Rozhen National Astronomical Observatory, which is operated by the Institute of Astronomy, Bulgarian Academy of Sciences and the 90-cm telescope of the University Observatory Jena, which is operated by the Astrophysical Institute of the Friedrich Schiller University.}}
\author[G. Maciejewski et al.]
{G.~Maciejewski$^{1,2}$\thanks{E-mail: gm@astro.uni-jena.de}, 
D.~Dimitrov$^{3}$, 
R.~Neuh\"auser$^{1}$, 
A.~Niedzielski$^{2}$, 
St. Raetz$^{1}$, 
\newauthor
Ch. Ginski$^{1}$, 
Ch.~Adam$^{1}$, 
C.~Marka$^{1}$, 
M.~Moualla$^{1}$ 
and M.~Mugrauer$^{1}$
\\
$^{1}$Astrophysikalisches Institut und Universit\"ats-Sternwarte, Schillerg\"asschen 2--3, D--07745 Jena, Germany\\
$^{2}$Toru\'n Centre for Astronomy, N. Copernicus University, Gagarina 11, PL--87100 Toru\'n, Poland\\
$^{3}$Institute of Astronomy, Bulgarian Academy of Sciences, 72 Tsarigradsko Chausse Blvd., 1784 Sofia, Bulgaria}
\begin{document}

\date{Accepted . Received ; in original form }

\pagerange{\pageref{firstpage}--\pageref{lastpage}} \pubyear{2010}

\maketitle

\label{firstpage}

\begin{abstract}
Photometric follow-ups of transiting exoplanets may lead to discoveries of additional, less massive bodies in extrasolar systems. This is possible by detecting and then analysing variations in transit timing of transiting exoplanets. We present photometric observations gathered in 2009 and 2010 for exoplanet WASP-3b during the dedicated transit-timing-variation campaign. The observed transit timing cannot be explained by a constant period but by a periodic variation in the observations minus calculations diagram. Simplified models assuming the existence of a perturbing planet in the system and reproducing the observed variations of timing residuals were identified by three-body simulations. We found that the configuration with the hypothetical second planet of the mass of $\sim$15 $M_{\earth}$, located close to the outer 2:1 mean motion resonance is the most likely scenario reproducing observed transit timing. We emphasize, however, that more observations are required to constrain better the parameters of the hypothetical second planet in WASP-3 system. For final interpretation not only transit timing but also photometric observations of the transit of the predicted second planet and the high precision radial-velocity data are needed.  
\end{abstract}


\section{Introduction}

%_TABLE 1_______________________________________________________
\begin{table*}
\centering
\begin{minipage}{115mm}
\caption{The summary of observing runs: $X$ -- the airmass range, $N_{\rmn{exp}}$ -- the number of useful exposures, $T_{\rmn{exp}}$ -- used exposure times, $FWHM$ - the averaged full width at half maximum of a stellar profile. Dates are given at the beginning of nights.} 
\label{tabela1}
\begin{tabular}{c l c c c c c}
\hline
Run & Date & Observatory & $X$ & $N_{\rmn{exp}}$ & $T_{\rmn{exp}}$ (s) & $FWHM$ ('') \\
\hline 
1 & 2009 July 28 & Rozhen & $1.01-1.29$ & $283$    & 30                 & $1.5$ \\
2 & 2009 Aug. 21 & Rozhen & $1.04-1.92$ & $200$    & 60                 & $1.5$ \\
3 & 2009 Sep. 03 & Rozhen & $1.01-1.38$ & $298$    & 30                 & $1.3$ \\
4 & 2009 Sep. 27 & Jena   & $1.05-1.94$ & $797$    & 8                  & $3.2$ \\
5 & 2009 Nov. 03 & Jena   & $1.13-2.32$ & $531$    & 10, 15             & $3.9$ \\
6 & 2010 Apr. 18 & Jena   & $1.08-2.19$ & $412$    & 25, 30             & $4.8$ \\
\hline
\end{tabular}
\end{minipage}
\end{table*}
%_____________________________________________________________

The analysis of observations minus calculations ($O-C$) diagrams is a great tool commonly used for astrophysical studies of eclipsing binaries or pulsating variables. This method has already brought discoveries of exoplanets. Variations in the $O-C$ diagram of the pulsating subdwarf B HS~2201+2610 were interpreted as a fingerprint of a giant planet \citep{silvottietal07}. More recently, two planets orbiting an eclipsing system HW~Vir were announced by \citet{leeetal09} and \citet{qianetal10} found a giant planet around the hibernating cataclysmic binary system QS~Vir. These discoveries are based on interpretation of the light-travel time effect generated by an additional body. 

The timing of known transiting planets is also expected to lead to discovering additional planets \citep{miralda02,holmanmurray05, agoletal05, steffenetal07}. In a single-planet extrasolar system a planet orbits its host star on a Keplerian orbit. If one assumes that the inclination of its orbit plane is close to $90\degr$, transits occur at equal intervals. If there is another (not necessarily transiting) planet in the system it interacts gravitationally with the transiting planet and generates deviations from the strictly Keplerian case. These perturbations result in a quasi-periodic signal in an $O-C$ diagram of the transiting planet. The transit timing variation (TTV) method can be sensitive to small perturbing masses in orbits near low-order mean-motion resonances (MMRs, \citealt{steffenetal07}). A terrestrial-mass planet perturbing a hot-Jupiter gas giant is expected to cause a TTV amplitude of $\sim$1 minute and this signal grows sharply as bodies approach a MMR \citep{steffenetal07}. One must note that deriving the orbital elements and mass of the perturber from TTV, however, is a difficult inverse problem (e.g. \citealt{nesvornymorbidelli08}). For a given transiting planet the TTV signal depends on 7 unknown parameters of the perturbing planet, i.e. its mass, semi-major axis, eccentricity, inclination, nodal and periapse longitudes and difference of mean orbital phases of both planets for a given epoch. Exploring such a 7-dimensional space of parameters is not a trivial task \citep{nesvornybeauge09}. The different configurations may generate similar TTV signals with the identical dominant periodicity \citep{fordholman07}.

The timing effects have been already studied in several transiting planet system: TrES-1 \citep{steffenagol05, rabusetal09}, HD~189733 \citep{milleretal08b,hrudkova10}, HD~209458 \citep{agolsteffen07, milleretal08a}, GJ~436 \citep{beanseifahrt08}, CoRoT-1 \citep{bean09}, TrES-2 \citep{rabusetal09}, TrES-3 \citep{gibsonetal09a}, and HAT-P-3 \citep{gibsonetal09b}. In all these cases the authors could only put constraints on the parameters of a hypothetical second planet because no significant signals in $O-C$ diagrams have been found so far. \citet{diazetal08} announced detection of period variation in OGLE-TR-111b and suggested that it may be caused by a perturbing Earth-mass planet in an outer orbit. However, \citet{adamsetal10} observed 6 new transits of OGLE-TR-111b and concluded that there is no compelling evidence for the timing variation in the system and placed an upper limit of about 1 $M_{\earth}$ on the mass of a potential second planet in a 2:1 orbital resonance.

Our first efforts to detect TTV signal for the transiting planets XO-1b, TrES-1b and TrES-2b resulted in redetermining their transit ephemerides \citep{raetzetal09a,raetzetal09b}. In 2009 we launched an international observing campaign to detect and characterise a TTV signal in selected transiting exoplanets. The programme is realised by collecting data from 0.6--2.2-m telescopes spread worldwide at different longitudes. In this paper we present photometric observations obtained for transiting exoplanet WASP-3b. 

The existence of a planet around the unevolved main sequence star WASP-3 ($V=10.485$ mag) was discovered by detecting transits followed by radial velocity measurements \citep{pollaccoetal08}. The exoplanet turned out to be a strongly irradiated gas giant. Its mass and radius were found to be $1.76^{+0.08}_{-0.14}$ $M_{\rmn{J}}$ and  $1.31^{+0.07}_{-0.14}$ $R_{\rmn{J}}$, respectively. It orbits its host star in $\sim$44 h in a circular orbit with the semi-major axis of $0.0317^{+0.0005}_{-0.0010}$ au. The orbital inclination was found to be $84.4^{+2.1}_{-0.8}$ deg. The host star has a spectral type of F7--8V and photospheric temperature of $6400\pm100$ K. \citet{gibsonetal08} observed two transits with a photometric precision of $\sim$4 millimag (mmag) and redetermined the planetary radius $1.29^{+0.05}_{-0.12}$ $R_{\rmn{J}}$ and the orbital inclination $85.06^{+0.16}_{-0.15}$ deg. The mean density of the planet was found to be $0.82^{+0.14}_{-0.09}$ $\rho_{\rmn{J}}$. Moreover, \citet{gibsonetal08} calculated new ephemeris which resulted in the orbital period of $1.846835\pm0.000002$ d. 

\citet{simpsonetal10} presented a spectroscopic observation of the Rossiter-McLaughlin effect for WASP-3 system. The sky-projected angle between the stellar rotation axis and planetary orbital axis $\lambda$ was found to be $15^{+10}_{-9}$ deg, i.e. consistent with zero within 2$\sigma$. More recently, \citet{tripathietal10} obtained spectra during two separate transits and got a more precise value $\lambda=3.3^{+2.5}_{-4.4}$ deg. These findings indicate that WASP-3b underwent relatively non-violent migration process which did not perturb it from the primordial alignment of the proto-planetary disk \citep{simpsonetal10}. Furthermore, \citet{tripathietal10} presented photometric observations of 6 transits, including one incomplete. Some deviations from the linear ephemeris were found and explained as a results of either genuine period variations or underestimating the transit time uncertainties. \citet{tripathietal10} also redetermined planets's mass which was found to be $2.04\pm0.07$ $M_{\rmn{J}}$.

The order of this paper is as follows. In Section 2 observations and data reduction are presented. The finding that the observed transit timing cannot be explained by a constant period is shown in Section 3. Preliminary configurations including a third body in the system (a perturbing planet), which may reproduce the observed variations of timing residuals, are identified and discussed in Section 4. Final conclusions are collected in Section 5.

% FIGURE 1
\begin{figure*}
  \includegraphics[width=17cm]{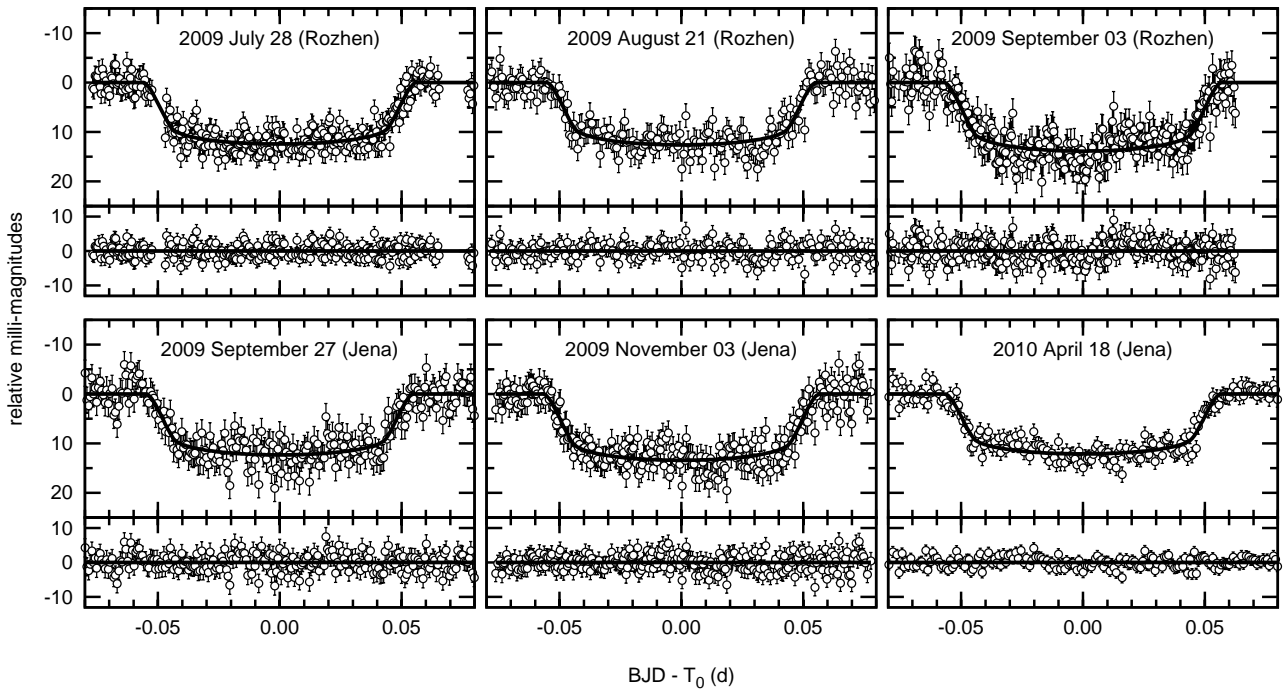}
  \caption{Light curves of WASP-3b transits in individual runs. The best-fitting models are shown as continuous lines. The residuals are presented in panels below each light curve.}
  \label{rys1}
\end{figure*}

\section[]{Observations and data reduction}

We observed 6 complete transits of WASP-3b with two telescopes both with the effective mirror diameter being 60 cm. While preparing observing runs only predicted complete transits during darkness were considered. Observations started $\sim$1 hour before the expected beginning of a transit and ended $\sim$1 hour after the event. Collecting long series of out-of-transit data is crucial in removing systematic trends and normalizing a light curve (e.g. \citealt{gibsonetal09a,hrudkova10}). All observations were acquired in the $R$-band filter \citep{bessel90}.

\subsection[]{Rozhen}

The CCD photometry of 3 transits on 2009 July 28, August 21 and September 3 was gathered with the 60-cm Cassegrain telescope at the Rozhen National Astronomical Observatory (NAO, Bulgaria), operated by the Institute of Astronomy, Bulgarian Academy of Sciences. The telescope was equipped with the CCD camera FLI PL09000 ($3056 \times 3056$, 12$\mu$m pixel). The field of view was $17\farcm3 \times 17\farcm3$ with the scale of $0.334$ arcsec per pixel. As a result of $3 \times 3$ binning the final scale was $1.017$ arcsec per pixel. Exposure times were selected on nightly basis to maximize the signal-to-noise (S/N) ratio, keeping pixels unsaturated and intensity in the linear region of the CCD ($<$50000 counts) for the object and probable standard stars.

The standard IDL procedures (adapted from DAOPHOT) were used for the reduction of the photometric data (dark frame subtraction and flat-fielding), and computing the differential aperture photometry. To measure instrumental magnitudes, we used apertures with radii of 3.2, 4, 5, 6, 7, 8 and 9 binned pixels and a background ring located between 12 and 16 pixels. The best precision was reached using an aperture of 4 binned pixels. Using the method of \citet{everetthowell01} several stars (4 to 6) with photometric precision better than 5 mmag were selected to create an artificial standard star used for differential photometry.

\subsection[]{University Observatory Jena}

Transits on 2009 September 27, November 3 and 2010 April 18 were monitored with the 90-cm reflector telescope at the University Observatory Jena. Observations were taken with the new CCD-imager STK (see \citealt{mugrauerberthold10}) which was installed in the Schmidt-focus of the telescope ($D=60$ cm, $f/D=3$). The camera exhibits $52\farcm8 \times 52\farcm8$ field of view with a pixel-scale of $1.546$ arcsec per pixel. The exposure time was set to get the highest S/N ratio for the target and was adjusted during a run due to changes of the airmass and atmospheric transparency. During third run the telescope was significantly defocused and hence longer exposure times were used. Applying this method was expected to minimise random and flat-fielding errors (e.g. \citealt{southworth09}), as confirmed by our higher-quality light curve (see bottom right panel in Fig.~\ref{rys1}).   

CCD frames were processed using a standard procedure that included subtraction of a dark frame, flat-fielding with twilight flats, differential aperture photometry, and astrometric calibration with the software pipeline developed for the Semi-Automatic Variability Search sky survey \citep*{niedzielskietal03}. The instrumental coordinates of stars in frames were transformed into equatorial ones making use of positions of stars brighter than $V=15.5$ mag and extracted from the Guide Star Catalogue. To measure instrumental magnitudes, we used apertures with radii of 5, 6, 7 and 8 pixels and a background ring located between 12 and 20 pixels. The 6 pixel aperture was found to produce light curves with the smallest scatter and they were used in further analysis. To generate an artificial comparison star, 30 per cent of the stars with the lowest light-curve scatter were selected iteratively from field stars not fainter than 3 magnitudes below the saturation level. To achieve the similar probing rate in all runs ($\sim$1 point per minute), 3 and 2 point binning of a final light curve was applied. The detailed summary of observing runs is presented in Table~\ref{tabela1}.

%_TABLE 2_______________________________________________________
\begin{table*}
\centering
\begin{minipage}{135mm}
\caption{Parameters of transit light-curve modelling. $T_{0}$, $T_{\rmn{d}}$, $\delta$, $\sigma$, and $E$ denote the mid-transit time, the transit time duration, the depth, the averaged standard deviation of the fit, and the epoch, respectively. The $O-C$ values are given both in days and in errors of mid-transit times. Note that BJD times are based on Terrestrial Dynamic Time (TT).} 
\label{tabela2}
\begin{tabular}{c c c c c c c c}
\hline
Run & $T_{0}$ & $T_{\rmn{d}}$ & $\delta$ & $\sigma$ & $E$ & $O-C$ &  $O-C$ \\
    & $\rmn{BJD}-2455000$  & (min)   & (mmag)   & (mmag)   &     & (d) & ($T_{0}$ errors) \\
\hline 
1 & $41.41271\pm0.00049$ & $161.0\pm1.6$ & $12.4\pm0.3$ & 1.6 & 236 & $-0.00046$ & $-0.9$ \\
2 & $65.41995\pm0.00059$ & $158.9\pm1.9$ & $12.6\pm0.4$ & 1.9 & 249 & $-0.00207$ & $-3.5$ \\
3 & $78.34873\pm0.00058$ & $163.7\pm1.9$ & $13.9\pm0.5$ & 2.4 & 256 & $-0.00114$ & $-2.0$ \\
4 & $102.35933\pm0.00056$& $157.9\pm1.9$ & $12.3\pm0.4$ & 2.1 & 269 & $+0.00120$ & $+2.1$ \\
5 & $139.29713\pm0.00049$& $162.9\pm1.6$ & $13.4\pm0.4$ & 1.7 & 289 & $+0.00169$ & $+3.4$ \\
6 & $305.51082\pm0.00039$& $162.6\pm1.3$ & $12.1\pm0.6$ & 1.2 & 379 & $+0.00018$ & $+0.5$ \\
\hline
\end{tabular}
\end{minipage}
\end{table*}
%_____________________________________________________________

\section[]{Results}
 
\subsection[]{Light curve analysis} 

A model-fitting algorithm available via the Exoplanet Transit Database \citep*{poddanyetal10} was used to derive transit parameters: transit duration, depth and mid-transit time and their errors. The procedure employs the \textsc{occultsmall} routine of \citet{mandelalgol02} and the Levenberg--Marquardt non-linear least squares fitting algorithm. The latter routine also provides errors for the fitted parameters (O. Pejcha 2010, private communication). Our tests showed that the errors of mid-transit times derived from Levenberg--Marquardt method are consistent within $\pm$30 per cent with these values obtained from a Markov Chain Monte Carlo algorithm. An impact parameter $b=a \cos i / R_{*} = 0.448\pm0.014$, where $a$ is semi-major axis, $i$ is inclination, and $R_{*}$ is host-star radius, was taken from \citet{gibsonetal08} and was fixed during the fitting procedure. The model-fitting algorithm uses the linear limb-darkening law, thus we used a linear limb-darkening law by \citet{vanhamme93} with the linear limb-darkening coefficient linearly interpolated for the host star. To determine the zero-point shift of magnitudes and to remove systematic trends which may exist in our data, a first- or second-order polynomial was used. The mid-transit times were corrected from UTC to Terrestrial Dynamic Time (TT) and then transformed into BJD. Light curves with best-fitting models and residuals are shown in Fig.~\ref{rys1}. Derived parameters are collected in Table~\ref{tabela2}.

We achieved an averaged photometric precision between 1.2 and 2.4 mmag. The mid-transit timing errors are in the range of 34--51 s. The mean transit duration was found to be $161.2\pm2.3$ min -- a value similar to $159.8^{+1.3}_{-2.6}$ min reported by \citet{pollaccoetal08} and noticeably smaller than $165.2^{+1.2}_{-0.8}$ min reported by \citet{gibsonetal08} and $168.8\pm0.7$ min announced by \citet{tripathietal10}. The mean transit depth was found to be $12.8\pm0.7$ mmag which results in a planet-to-star radii ratio $\rho=0.108\pm0.003$. This value is within the range of planet-to-star radii ratios determined for individual transits by \citet{tripathietal10}. \citet{pollaccoetal08} and \citet{gibsonetal08} obtained smaller values, i.e. $\rho=0.1030^{+0.0010}_{-0.0015}$ and $\rho=0.1014^{+0.0010}_{-0.0008}$, respectively.

\subsection[]{Transit ephemeris} 

Having a long time span of observations we determined a new ephemeris. As a result of fitting a linear function of epoch and period $P$, we obtained:
\[
	T_{0}  = 2454605.56000 \pm 0.00011 \,  \; \rmn{(BJD, based~on~TT)} 
\]
\[
	P      = 1.8468355 \pm 0.0000007 \,  \; \rmn{d}. 
\]
Individual mid-transit errors were taken as weights. 

\begin{figure}
 \includegraphics[width=84mm]{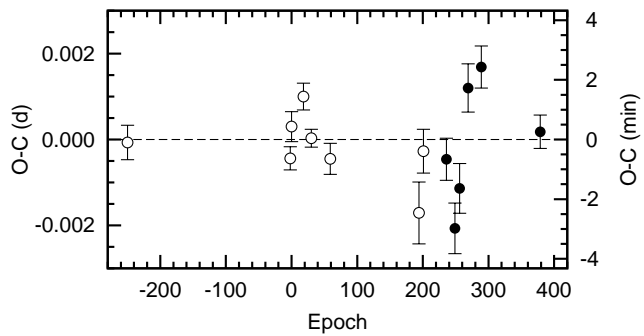}
 \caption{The observed minus calculated diagram for WASP-3b. The open symbols denote literature data (complete transits only) taken from \citet{pollaccoetal08}, \citet{gibsonetal08} and \citet{tripathietal10} while the filled ones denote results reported in this paper. Timing residuals were calculated using a new ephemeris derived in this study.}
 \label{rys2}
\end{figure}

The $O-C$ diagram was generated using the new ephemeris. The timing residuals are collected in Table~\ref{tabela2}. The $O-C$ diagram is plotted in Fig.~\ref{rys2} where besides transit times reported in this paper the literature data are also shown. Following \citet{adamsetal10} we have confirmed that the times published by \citet{pollaccoetal08}, \citet{gibsonetal08} and \citet{tripathietal10} do not account for the UTC--TT correction (D. Pollacco 2010, private communication; A. Tripathi, private communication). Therefore, we added appropriate corrections to published times before analysing the $O-C$ diagram. In the case of 3 points (1 from \citealt{tripathietal10} and 2 reported in this paper) timing residuals have a significance greater than 3 $\sigma$. That indicates that the uncertainties are underestimated or that the orbital period is not constant. 

It is worth mentioning that data points from Rozhen lie below or close to zero level in the $O-C$ diagram while those from Jena lie above it. This finding could suggest the presence of the systematic offset between both telescopes. However, our experience shows that this scenario seems to be unlikely. Simultaneous observations collected by both telescopes for another transiting planet gave difference between mid-transit times of 10 s only -- much less than the error bars (Maciejewski et al. 2010, in prep.). One must note that error underestimating cannot be completely ruled out in our determinations but it would have a negligible influence on data point distribution and final conclusions.  
       
% FIGURE 2

\subsection[]{Search for a cyclic TTV} 

\begin{figure}
 \includegraphics[width=8.4cm]{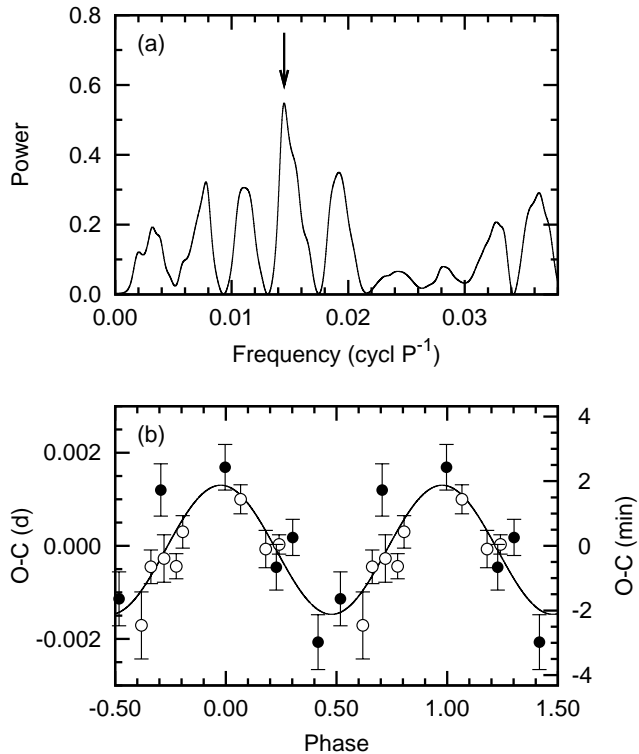}
 \caption{(a) A periodogram generated for timing residuals plotted in Fig~\ref{rys2}, showing a peak at $f_{\rmn{ttv}}=0.0145$ cycl $P^{-1}$. (b) The $O-C$ diagram phased with $P_{\rm{ttv}}=69.0\pm2.4$ $P$ corresponding to $f_{\rmn{ttv}}$. The open and filled symbols denote literature and this paper (see Table~\ref{tabela2}) data, respectively. Note that data points were duplicated to cover phases from $-0.5$ to $1.5$.}
 \label{rys3}
\end{figure}

The timing residuals were searched for cyclic transit time variation by means of the analysis of variance method (ANOVA, \citealt{schwarzenberg96}) to identify any significant periodicities. Fig.~\ref{rys3}a presents the generated periodogram covering frequencies $f$ in the range of 0.000--0.038 cycl $P^{-1}$ where the upper limit is the calculated Nyquist frequency $f_{\rmn{N}}$. A significant peak was found around $f_{\rmn{ttv}}=0.0145$ cycl $P^{-1}$ which corresponds to a period of $P_{\rm{ttv}}=69.0\pm2.4$ $P$. The $O-C$ diagram phase-folded at $P_{\rm{ttv}}$ is plotted in Fig.~\ref{rys3}b where the best-fitting sinusoid is also sketched. The semi-amplitude of variation was found to be $0.0014\pm0.0002$ d.

% FIGURE 3

\section[]{Discussion}
 
\subsection[]{Exomoon} 

A sinusoidal TTV might be caused by an exomoon around the transiting planet \citep*{simon07,kipping09}. \citet{kipping09} predicts that an exomoon also produces transit duration variation (TDV) with the same period as TTV and shifted by $\pi/2$ in phase. That effect is not visible in TDV of WASP-3b. If the hypothetical moon is located at a Hill radius away from the planet, its mass is expected to be 6 per cent of the planetary mass to produce the observed TTV amplitude. This extreme value is a lower limit of the mass because the needed mass increases as the moon is closer to the planet. Both tests indicate that the exomoon hypothesis is unlikely here. 

\subsection[]{Reanalysis of the radial velocity data} 

The possible non-zero eccentricity of the transiting planet may be a results of confusion with a two-planet system in which both planets orbit its host star in circular orbits and stay in an inner 2:1 resonance \citep{angladaetal10}. This degeneracy is a consequence of the Fourier expansion of the Kepler equation into powers of eccentricity. The non-zero eccentricity of the transiting planet may also betray the presence of an outer perturber in the system (see e.g. \citealt{fabrycky09}). Tidal interactions with the star are expected to circularise the orbit of the transiting planet \citep{zahn77} but the presence of the second planet in the wider orbit may significantly delay this process \citep{mardling07}.  

The eccentricity of WASP-3b was reported to be statistically indistinguishable from zero. \citet{pollaccoetal08} obtained $e=0.05\pm0.05$ and \citet{simpsonetal10} got $e=0.07\pm0.08$. Authors of both papers adopted $e=0.0$ in their analysis. \citet{tripathietal10} adopted a priori circular orbit. We reanalysed the joined radial velocity data sets from \citet{pollaccoetal08} and \citet{tripathietal10} (only out-of-transit measurements) to test solutions with WASP-3b orbiting in an eccentric orbit. We used the \textsc{Systemic Console} software \citep{meschiarietal09}. The zero eccentricity Keplerian fit gives a reduced $\chi^2=3.3$ with $rms = 15.4$ m s$^{-1}$. If the eccentricity $e$ is set as a free parameter the best-fitting model results in $e=0.05\pm0.04$ with a lower reduced $\chi^2=2.3$ and $rms = 13.9$ m s$^{-1}$. Our results show that the orbit of WASP-3b may be circular within 2$\sigma$ and the eccentric orbit cannot be excluded. The remaining dispersion of the radial velocity is typical for fast rotating late-type F stars \citep*{saar98}.

\subsection[]{Inner perturber} 

The mass of the perturbing planet in an inner 2:1 resonance, $M_{\rmn{p}}$, depends on the mass of the outer, more massive transiting planet $M_{\rmn{t}}$ and its apparent eccentricity $e$ as follows:
\begin{equation}
	M_{\rmn{p}} = \frac{e}{\sqrt[3]{2}} M_{\rmn{t}}
	\label{rownanie1}
\end{equation}
\citep{angladaetal10}. In this case $M_{\rmn{p}}$ should be treated as a upper limit because it was obtained assuming circular orbits. The hypothetical inner, less massive planet is expected to produce perturbations which should be visible in TTV of WASP-3b. 

To check if an inner perturber in the WASP-3 system may reproduce observed cyclic TTV, we generated 2232 synthetic $O-C$ diagrams for orbits close to the 1:2 resonance. Following \citet{rabusetal09} we neglected the light-time effects and assumed that the planetary orbits are coplanar. We used the \textsc{Mercury} package \citep{chambers99} and the Bulirsch--Stoer algorithm to integrate the equations of motion for this three-body problem. Parameters of the transiting planet and its host star were taken from \citet{tripathietal10} and \citet{pollaccoetal08}, respectively. The initial circular orbits were assumed. The semi-major axis of the perturber $a_{\rmn{p}}$ was between 0.0185 and 0.0215 au ($\pm0.0015$ au away from the 2:1 resonance) in steps of $0.0001$ au. The mass of the second planet $M_{\rmn{p}}$ was varied from 5 to 30 $M_{\earth}$ in steps of 5 $M_{\earth}$. The upper limit was calculated according to equation~(\ref{rownanie1}) assuming $e=0.05\pm0.04$. The initial phase shift toward the transiting planet (in fact, an initial mean anomaly of the perturber), $\phi$, was between $0\degr$ and $330\degr$ in steps of $30\degr$. Computations covered 640 periods of WASP-3b, i.e. the time span covered by observations. Synthetic $O-C$ diagrams were searched for periodicity with the ANOVA method and the mean amplitude of variation was determined. The procedure was repeated for the 3:2 resonance.

Periodicity close to $P_{\rm{ttv}}$ was not found near either resonance. That allowed us to eliminate the scenario where the inner perturber generates the observed TTV signal.

%_TABLE 3_______________________________________________________
\begin{table}
\centering
\begin{minipage}{84mm}
\caption{Outer-perturber solutions which reproduce the observed $O-C$ variation. $a_{\rmn{p}}$ denotes the semi-major axis of the perturbing planet, $M_{\rmn{p}}$ is its mass, $P_{\rmn{p}}$ is its orbital period, $K_{\rmn{p}}$ is the expected semi-amplitude of the radial-velocity variation and $\chi^2_{\rm{red}}$ is the lowest value of reduced chi-square for direct model fitting.} 
\label{tabela3}
\begin{tabular}{c c c c c}
\hline
$a_{\rmn{p}}$ & $M_{\rmn{p}}$  & $P_{\rmn{p}}$ & $K_{\rmn{p}}$ & $\chi^2_{\rm{red}}$ \\
(au)          & $(M_{\earth})$ & (d)         & (m s$^{-1})$ \\
\hline 
0.0441 -- 0.0443 & 6 -- 10 & 3.03 -- 3.05 & 2.3 -- 3.8 & 2.4\\
0.0493 -- 0.0498 & 10      & 3.58 -- 3.64 & 3.7        & 2.8\\
0.0506 -- 0.0511 & 15      & 3.72 -- 3.78 & 5.3        & 1.5\\
\hline
\end{tabular}
\end{minipage}
\end{table}

\subsection[]{Outer perturber} 

Applying methods described in section 4.3, we generated 19680 synthetic $O-C$ diagrams assuming the existence of an outer perturber in the system. Both planets were assumed to orbit the star in initial circular orbits. We investigated cases with the second planet close to 2:1, 3:1, 3:2, and 5:3 MMRs, $\pm0.0020$ au away from each resonance. The mass of the perturber $M_{\rmn{p}}$ was varied from 5 to 50 $M_{\earth}$ in steps of 5 $M_{\earth}$. Solutions close to observed cyclic TTV were studied in detail by varying $M_{\rmn{p}}$ and $\phi$ in steps $\Delta M_{\rmn{p}} = 1$ $M_{\earth}$ and $\Delta \phi = 15\degr$, respectively.

\begin{figure}
 \includegraphics[width=8.4cm]{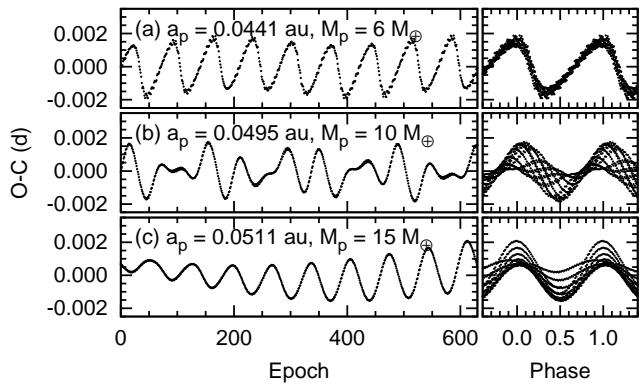}
 \caption{Examples of synthetic $O-C$ diagrams which reproduce the observed variations, generated for an outer perturber. Left panels -- the unfolded TTV signal, right panels -- the TTV signal folded to period close to $P_{\rm{ttv}}$.}
 \label{rys4}
\end{figure}

We found a family of solutions close to 5:3 MMR recovering the period close to $P_{\rm{ttv}}$ for $a_{\rmn{p}}$ between $0.0441$ and $0.0443$ au and $M_{\rmn{p}}$ between 7 and 10 $M_{\earth}$. The exemplary synthetic $O-C$ diagram is plotted in Fig.~\ref{rys4}a. We also found 2 groups of solutions close to the 2:1 MMR. In the first one the perturber has $M_{\rmn{p}}$ of 10 $M_{\earth}$ and $a_{\rmn{p}}$ between $0.0493$ and $0.0498$ au. Fig.~\ref{rys4}b shows the shape of expected signal which, despite being periodic, clearly shows a deformation of the amplitude. The second group of solutions may be found for $a_{\rmn{p}}$ between $0.0506$ and $0.0511$ au and $M_{\rmn{p}}$ of 15 $M_{\earth}$. As it is shown in Fig.~\ref{rys4}c, periodic variation may be modulated by a long-term periodicity significantly affecting the amplitude of variations. This long-term cyclic variability is caused by the resonant oscillations of the planetary orbits. Table~\ref{tabela3} summarises parameters of proposed models. 

No solutions close to observed TTV variation were found near resonances 3:1 nor 3:2. Orbits far away from MMRs were analysed using the \textsc{ptmet} code based on the perturbation theory \citep{nesvornymorbidelli08, nesvorny09}. A perturbing planet was put in initial circular orbits with semi-major axis $a_{\rmn{p}}$ between 0.0370 and 0.1200 au in steps of $0.0002$ au. As the amplitude of TTV signal was found to scale nearly linearly with the perturber's mass \citep{nesvornymorbidelli08}, $M_{\rmn{p}}$ was fixed and set equal 50 $M_{\earth}$. An initial phase shift toward the transiting planet $\phi$ was between $0\degr$ and $315\degr$ in steps $\Delta \phi=45\degr$. A total of 3320 systems were analysed but no configuration reproducing the observed variation was found. 

\begin{figure}
 \includegraphics[width=8.4cm]{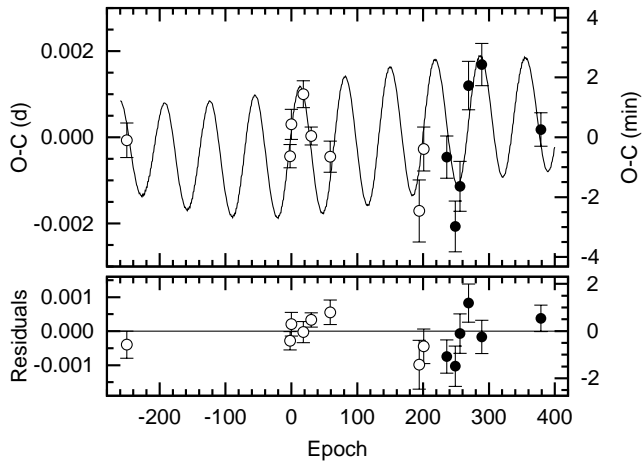}
 \caption{The $O-C$ diagram with the best-fitting model which reproduces the observed variations. The mass of the perturber is 15 $M_{\earth}$ and the semi-major axis of its initially circular orbit is 0.0507 au. Symbols are the same as in Fig.~\ref{rys2}. Residuals are plotted in the bottom panel. }
 \label{rys5}
\end{figure}

To quantify the quality of solutions collected in Table~\ref{tabela3}, the synthetic $O-C$ diagrams were directly fitted to data points. The chi-square test favours the third group of solutions where the smallest reduced $\chi^2_{\rm{red}} = 1.5$ was obtained for $a_{\rmn{p}}=0.0507$. This value is significantly smaller than $\chi^2_{\rm{red}} = 3.9$ for a model without a perturbing planet. The $O-C$ diagram with the best-fitting model is presented in Fig.~\ref{rys5} where the residuals are also plotted. The expected semi-amplitude of the radial velocity variation caused by the perturber is expected to be $\sim$5 m s$^{-1}$ and thus undetectable in the published radial velocity measurements. The most probable model was found to be stable dynamically. A simulation was run for $10^6$ yr and no significant changes in the eccentricities were detected. The eccentricities of transiting and perturbing planet were found to be below 0.005 and 0.033, respectively.

Such a WASP-3 system would not be completely unusual with two planets close to the 2:1 MMR (see e.g. Gliese 876 b and c, \citealt{marcyetal01}) but the first one with such close-in planets and the less massive planet being in an outer orbit. The origin of such a system would be interesting in a context of the large-scale orbital migration in young planetary systems. A Super-Earth (10 $M_{\earth}$) is expected to be caught in a trap when it is very close to the edge of the wide gap opened in a disk by a Jupiter-mass gas giant. This trapping prevents the outer planet from being locked in the 2:1 MMR. If a planet increases its mass by accretion, it may be released from such a trap and form a resonant configuration \citep{podlewska09}. This scenario cannot be rejected in the case of the outer perturber in the WASP-3 system because the nowadays planet might be a rocky remnant core of a planet which has greater mass in the past -- a hot Jupiter whose gaseous envelope was completely evaporated due to the proximity to the host star \citep{jacksonetal10}.

% FIGURE 4
%_____________________________________________________________

% FIGURE 5
%_____________________________________________________________

\section{Conclusions}

In this paper we show that transit timing of WASP-3b cannot be explained by a constant period of the exoplanet. We emphasise that this finding is based on only a few data points. However, one may put forward provisional hypothesis which assumes the existence of a perturber in the system. As a result of simplified $3$-body simulations, two groups of configurations which reproduce the observed TTV were identified close to the outer 2:1 MMR and one group close to the 5:3 MMR. A model with the hypothetical second planet of the mass of $\sim$15 $M_{\earth}$ and orbital semi-major axis of 0.0507 au was found to be most likely. In this scenario both planets stay in the stable orbital resonance with a period ratio of 2.02, i.e. very close to the 2:1 MMR. 

The expected radial-velocity semi-amplitude of the hypothetical second planet would be much smaller than the stellar jitter of WASP-3. This finding reduces chances of a direct detection of the perturber's signal by further radial-velocity follow-ups. However, we showed that one cannot exclude the non-zero eccentricity of WASP-3b. Such an eccentric orbital solution may be an indirect fingerprint of the second planet in a wider orbit gravitationally pumping the eccentricity of the transiting planet.

Assuming the hypothetical second planet is also a transiting exoplanet, it would cause periodic flux drops of the host star. The depth of these transits would be in the range of 0.03--0.35 per cent (or 0.3--3.8 mmag) depending on the adopted mean planetary density -- for a rocky planet (e.g. CoRoT-7b, \citealt{lageretal09,quelozetal09}) or an ultra-low-density hot Neptune (e.g. WASP-17b, \citealt{andersonetal10}), respectively. If one considers the mass of the hypothetical second planet and its proximity to the host star, the latter scenario seems to be less probable, causing transits to be shallow and difficult to detect. 

As the TTV method requires many high-quality light curves, more data will be gathered for WASP-3b to confirm or disprove the claimed variation in the $O-C$ diagram. A long-time baseline of observations is needed to characterise possible resonant oscillations in the TTV signal. More accurate radial-velocity measurements are also required to gain deeper insight into orbital parameters of the transiting planet. Additionally, a transit of the predicted second planet could be observable with a large ground-based or space-based telescope.

\section*{Acknowledgments}

GM and SR acknowledge support from the EU in the FP6 MC ToK project MTKD-CT-2006-042514. SR would also like to acknowledge support from DFG in program NE 515/32-1. DD acknowledges support from the project DO 02-362 of the Bulgarian Scientific Foundation. RN would like to acknowledge general support from DFG in programmes NE 515/23-1, 30-1 and 32-1. GM, RN and AN acknowledge support from the DAAD PPP/MNiSW project 50724260 \textit{Eclipsing binaries in young clusters and planet transit time variations}. AN was also supported by the Polish Ministry of Science and Higher Education grant NN203 510938. CG and CM would like to thank the German National Science Foundation (DFG) for support in programs NE 515/30-1 and SCHR 665/7-1, respectively. M.Moualla would like to thank the Syrian government for their support. Part of this paper is a result of PAN/BAN exchange and joint research project \textit{Spectral and photometric studies of variable stars}. This work has used Exoplanet Transit Database, http://var.astro.cz/ETD. We thank A.V.~Krivov for helpful comments on an earlier version of the paper. Finally, we owe a debt of gratitude to the anonymous referee for constructive remarks and suggestions which improved the paper.

\bsp

\label{lastpage}

\end{document}